# In-beam test results of an RPC-based module for position-sensitive neutron detectors with timing readout

G. Canezin[a], L. M. S. Margato[a], A. Morozov[a, 1], A. Blanco[a], J. Saraiva[a], L. Lopes[a], P. Fonte[a,b], Chung Chuan Lai[c], Per-Olof Svensson[c], G. Markaj[d], Florian M. Piegsa[d]

[a] *LIP-Coimbra, Departamento de Física, Universidade de Coimbra, Rua Larga, 3004-516 Coimbra, Portugal*

[b] *Coimbra Polytechnic – ISEC, Coimbra, Portugal*

[c] *Detector Group, European Spallation Source ERIC (ESS), Partikelgatan 2, 224 84 Lund, Sweden*

[d] *Laboratory for High Energy Physics, Albert Einstein Center for Fundamental Physics, University of Bern, Switzerland*

**Abstract**

Recently we have proposed a new concept of a thermal neutron detector based on resistive plate chambers and $^{10}B_4C$ solid neutron converters, enabling to readout with high resolution in both the 3D position of neutron capture and the neutron time of flight (ToF). In this paper, we report the results of the first beam tests conducted with a new neutron RPC detection module, coupled to the position readout units of a new design. The main focus is on the measurements of the neutron ToF and identification of the converter layer where the neutron is captured, giving the position along the beam direction.

## 1. Introduction

LIP-Coimbra laboratory is developing a neutron detection technology based on Resistive Plate Chambers (RPC) coated with $^{10}B_4C$ solid neutron converters [1-6]. This work is conducted in collaboration with the Detector Coatings Workshop of the European Spallation Source (ESS), which develops and manufactures the neutron converter coatings [7-9].

The feasibility of the $^{10}B$-RPC technology for building high-efficiency neutron detectors with sub-millimeter spatial resolution has already been demonstrated [3,5]. Compared to the other neutron detection technologies not based on the $^3He$, it also offers an attractive combination of high modularity, scalability to large areas, and low cost per unit area [1].

The next phase in this development is to explore the timing capability of the RPC-based thermal neutron detectors, which is expected to be their strength as it has already been demonstrated that RPCs can achieve sub-ns time resolution (see, e.g., [10]). In our recent paper, we proposed a new concept of a position-sensitive neutron detector with four-dimensional readout capability (XYZ and time) [6]. The detector consists of a stack of ten neutron RPC detection modules with the cathode plates lined on both sides with a $^{10}B_4C$ converter. The modules are positioned in alternative order with the position readout units of a new design. This type of detector can be suitable for the future generation of neutron science instruments at neutron facilities such as ESS, for applications that require high neutron detection efficiency combined with high spatial and timing resolution.

---

[1]Corresponding author: margato@coimbra.lip.pt

This paper reports the results of evaluation tests of a neutron RPC detection module from such a detector at the SINQ neutron spallation source at the Paul Scherrer Institut (PSI). The main focus is on the measurements of the neutron ToF and identification of the converter layer where the neutron is captured, giving the position along the beam direction.

## 2. Materials and methods

### 2.1. Detection module test setup

A single neutron RPC detection module (section 2.1.1), is sandwiched between two XY-position readout units (section 2.1.2). This ensemble is mounted inside a gas-tight aluminium enclosure with a 0.2 mm thick aluminium entrance window. This enclosure is filled with the R134a working gas ($C_2H_2F_4$) at atmospheric pressure, and a continuous gas flow of ~2 cc/min is maintained during operation.

#### 2.1.1. Neutron detection module

The detection module consists of a double-gap hybrid RPC composed of two resistive anode plates and an aluminium cathode coated on both sides with a thin film of $^{10}B_4C$. The sensitivity to thermal neutrons is achieved through the $^{10}B(n,^4He)^7Li$ capture reaction in this converter layer. The ionisation generated by the reaction products ($^4He$ or $^7Li$ ion) entering the gas gap leads to electron multiplication under a high electric field. Charge signals are induced on pickup electrodes by the drift of electrons and ions toward the anode and cathode, respectively.

A schematic drawing of the module is shown in Figure 1. The cathode, made of aluminium alloy 5754, has an area of 190 x 190 mm$^2$ and 0.3 mm thickness. It is coated on both sides with a 0.4 µm thick layer of $^{10}B_4C$ with a $^{10}B$ enrichment level of ~97%. The deposition of $^{10}B_4C$ was performed at the ESS Detector Coatings Workshop. The anodes are made of soda lime glass and have 200 x 200 mm$^2$ surface area and 0.33 mm thickness. The width of the gas gaps between the cathodes and the anodes is defined by 0.28 mm diameter spacers (PEEK monofilaments). The glass is painted on the side opposite to the gas gap with a resistive ink with a surface resistivity of $10^8$ Ω/□. The ink layer is ~0.1 mm thick and is used to uniformly distribute the electrical potential over the anode surface.

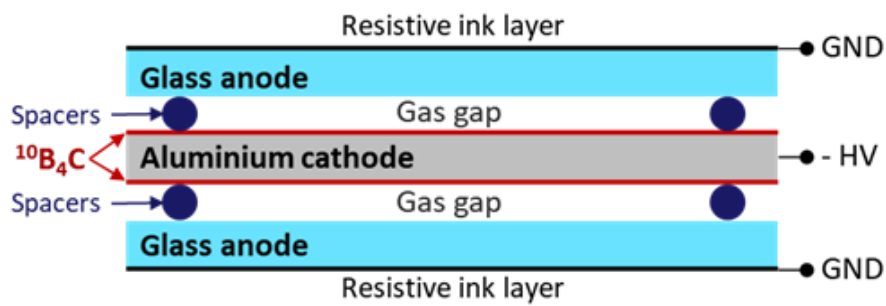

Figure 1: Schematic drawing of the detection module. Dimensions are not to scale.

## 2.1.2. Position and time readout

The XY-position readout unit is a two-layer flexible printed circuit board (FPCB) with two arrays of signal pickup electrodes (see Figure 2). The FPCB has a 25 μm thick polyimide film with 18 μm thick copper layers on both sides. The copper layers are etched to form arrays of 182 parallel strips. The strips are 0.3 mm wide and form a regular array with a pitch of 1 mm. The arrays are aligned orthogonally to each other and thus allow to readout X and Y coordinates of the detected neutrons by sensing the induced charge. An anisotropic electrically adhesive transfer tape type 9703 from 3M (ACF) is used to interconnect the strips between the FPCBs. This approach allows to simplify the assembly of the detector by eliminating the need to solder a large number of strips by hand.

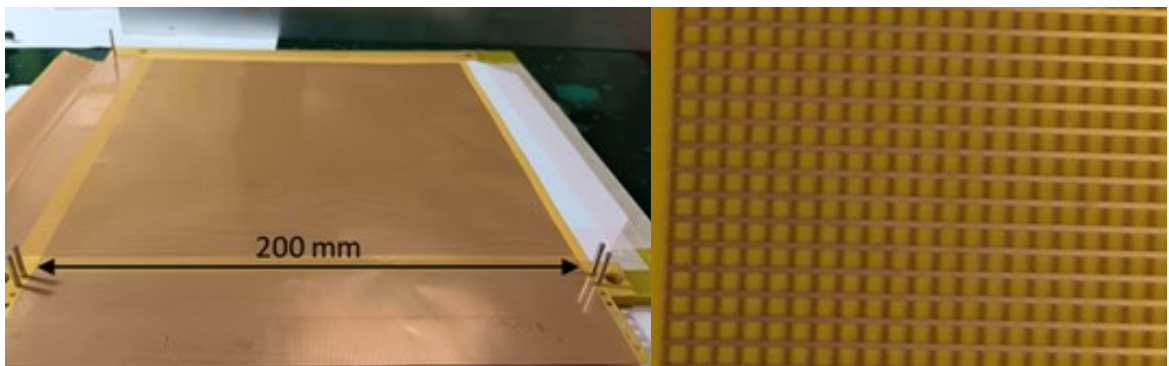

Figure 2. a) Photograph of an entire FPCB. b) Photograph of a side area of the FPCB showing two orthogonal arrays of copper strips (0.3 mm wide, 1 mm pitch). The array on the back sides of the FPCB is visible due to the high transparency of the thin polyimide film.

Two XY-position readout units are required to read one double-gap RPC (see Figure 3). For the X direction (the same also applies to Y), the strips with the same index in both units are interconnected and fed into the same charge preamplifier. This approach allows us to reduce by a factor of two the number of electronic channels needed to read the XY coordinates of the detected events. As discussed in [6], positioning the readout units in the stack in a way that maintains the X side orientation (e.g. on the bottom), the comparison of the sum signal in X and Y directions should allow to identify the gas gap (for example, above or below the cathode in Figure 3) which was triggered by a particular event. This is possible due to the fact that the array not facing the triggered gas gap is partially screened by the array which is directly in front of the gas gap [6].

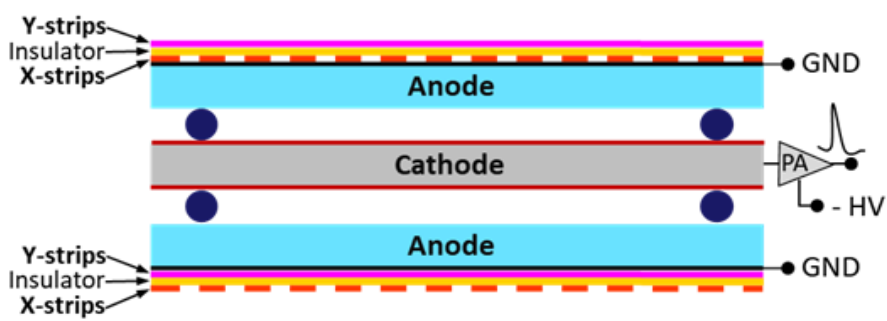

Figure 3: Schematic drawing illustrating two position readout units positioned above and below the detection module. Note that several detection modules and readout units can be stacked in an alternative order to achieve high detection efficiency.

The timestamp of a detected neutron event is readout using the signal induced in the cathode. The induced signal has two components: a fast one resulting from the drift of electrons in the gas gap, and a slow one originating from the ion drift. The fast component is used to define the trigger for the acquisition system, while the integral of the slow component is also recorded as this information can be useful for event classification in post-processing.

**2.2. Front end electronics and DAQ**

The XY signal pickup strips are read by custom-made charge sensitive preamplifiers (PA) developed at LIP [11]. The PAs have a gain of 250 mV/pC and an integration time constant of 20 µs. The cathode signal is fed into two filters, one which allows to pass only the slow component of the signal while the other only the fast one. The first filter is read by a PA of the same type as used for the strips, and the second by a timing amplifier that provides a Low Voltage Differential Signalling (LVDS) trigger when the signal amplitude exceeds a defined threshold.

Signals from the PAs are registered by a data acquisition system (DAQ) based on the Triggered Readout Board (TRB) family developed at GSI Institute [12]. The board is equipped with "ADC add-on" modules of the same family, each featuring four 48-channel 12-bit streaming ADCs with a 40 MHz sampling rate and in-board digital pulse processing. To record the waveforms from the PAs channels, the DAQ has been configured with a time window of 4 µs and a sampling time of 200 ns (20 samples per waveform).

The DAQ is triggered by the LVDS signal from the timing amplifier reading the cathode. The time stamp provided by these signals is also registered by the DAQ for each neutron event.

**2.3. Experimental setup**

During the experimental campaign at the Swiss spallation neutron source (SINQ) at Paul Scherrer Institute (PSI), the neutron detection module was tested at the NARZISS and BOA beamlines.

The preliminary tests were conducted at the NARZISS beamline with monochromatic neutrons of 4.96 Å. The beam was collimated by two slits placed at the distances of 2100 mm and 140 mm from the detector's entrance window. The slits had an aperture of 1 mm x 10 mm and the detector was oriented normally to the beam. The cathode signal was fed to an ORTEC preamplifier 142AH (gain of ~0.7 V/pC), followed by a linear amplifier. The amplitude spectrum of the signals at the output of the amplifier was recorded by a multichannel analyzer (MCA-8000D from Amptek).

The main tests were conducted at the BOA beamline. A schematic drawing of the experimental system is shown in Figure 4. The neutron beam is first collimated by a Boral aperture with a 10 mm x 10 mm aperture. Then the chopper shapes the quasi-continuous neutron beam into neutron pulses with a frequency of 19 Hz. The distance between the chopper and the detector is 6870 mm. An evacuated tube with sapphire windows on both sides is installed between the chopper and the detector to avoid scattering of neutrons in the air. A neutron absorber plate with a 25 mm x 10 mm aperture is placed directly in front of the detector window.

The neutron ToF is computed as the time difference between the two timestamps: the one given by the chopper and the one defined by the fast component of the RPC cathode signal. The obtained ToF was corrected for the delay between the chopper trigger signal and the moment of time when the chopper is fully open (0.26 ms).

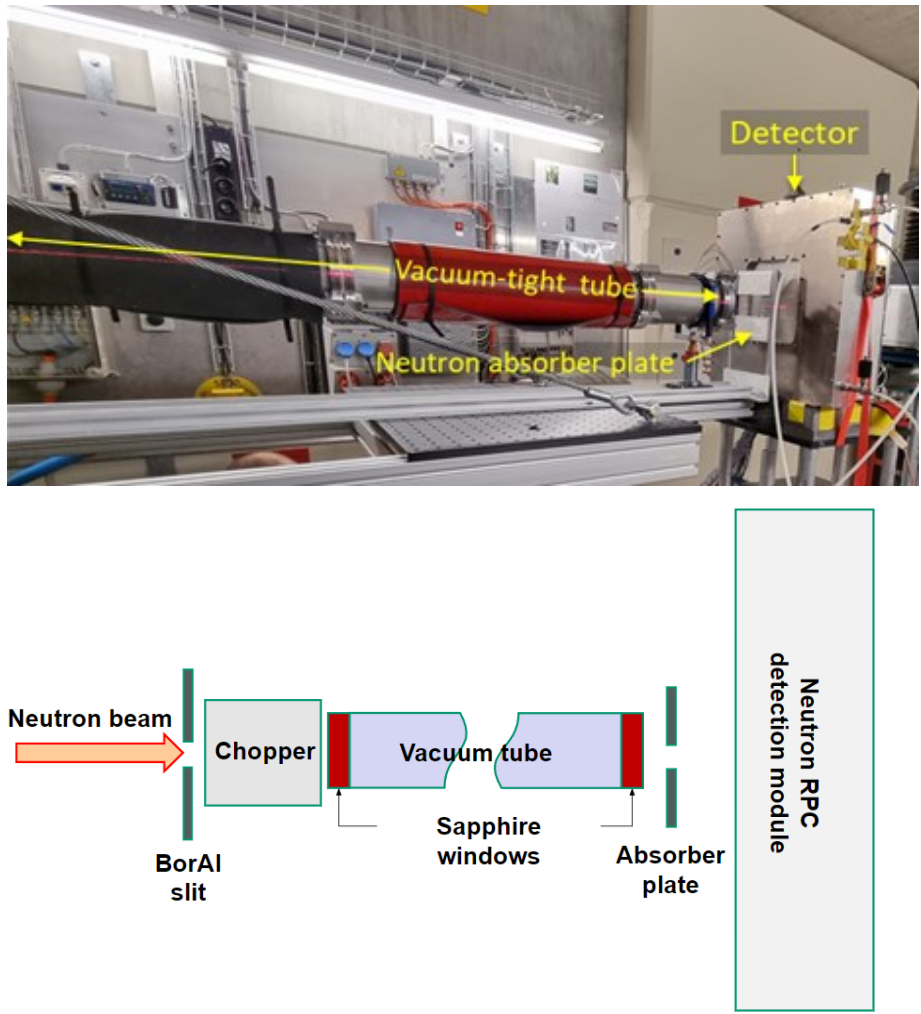

Figure 4. Photograph and the schematic drawing of the experimental setup at the BOA beamline.

### 2.4. Simulation of the detection efficiency

One of the objectives of this study was to obtain the wavelength spectrum of the beam at the BOA beamline based on the ToF data measured with our detector. This procedure requires knowledge of the dependence of the neutron detection efficiency on the neutron wavelength.

The efficiency was evaluated in Monte Carlo simulations. We have used Geant4 v10.7.2 with the QGSP_BIC_HP physics list [13] and ANTS2 v4.36 [14] as the front end. The simulations were conducted for ~20 neutron wavelengths, ranging from 0.1 Å to 20 Å. For each neutron wavelength, $10^6$ primary neutrons were generated, and, like in our previous work [4], a neutron event was considered detected when one of the products of the capture reaction deposited at least 100 keV in one of the RPC gas gaps. The simulation results are shown in Figure 5.

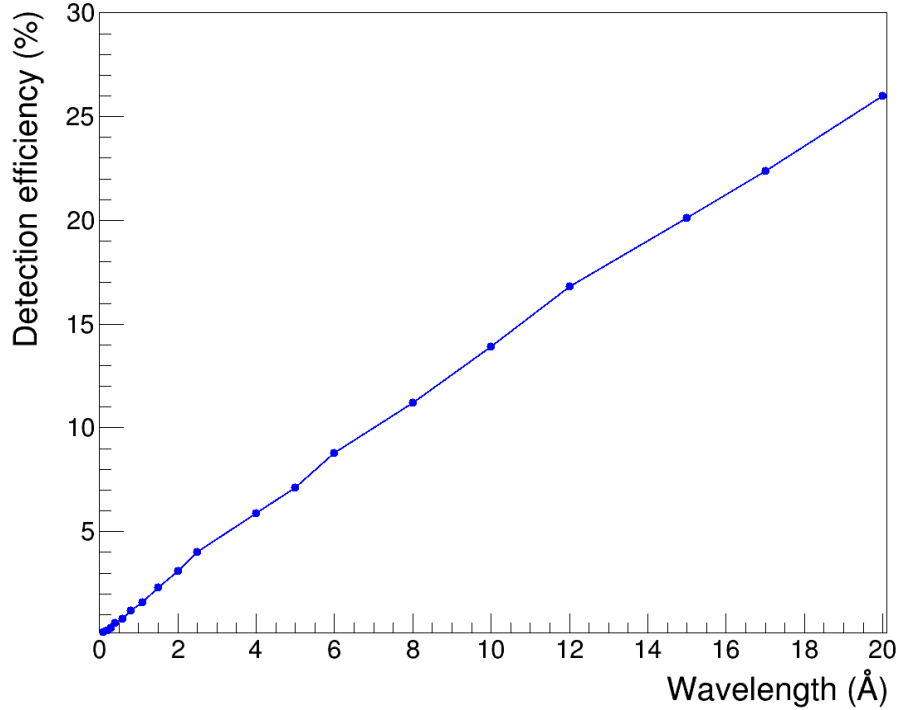

Figure 5. Detection efficiency as a function of the neutron wavelength for the detection module obtained in Monte Carlo simulations. The statistical uncertainties are smaller than the size of the markers and the line in the figure serves as a guide to the eye.

## 3. Results and discussion

### 3.1. Plateau curve of the detection module

The measured count rate as a function of the HV applied to the RPC (plateau curve) for the detection module irradiated by the neutron beam at the NARZISS beamline is shown in Figure 6. The curve shows a knee at 1800 V and the plateau extends for about 200 V before the onset of the exponential growth.

The results obtained here with a 0.28 mm gas gap significantly differ from the ones observed before with a 0.35 mm gas gap: the knee appeared at 2050 V and the plateau was extending for about 500 V [3]. The appearance of the knee at lower voltages in this study can be explained by the fact that for the same voltage applied to the RPC, the magnitude of the electric field in the gas gap is higher for a thinner gas gap. Besides that, on average, the energy deposited in the gas by the $^4$He and $^7$Li particles from the neutron capture reaction in a $^{10}B_4C$ layer, and the path for the electron multiplication, is lower than for wider gas gaps. Consequently, the amplitude of the induced signals starts to become comparable to that of minimum ionising particles (MIPs), such as the Compton electrons from gamma rays interaction. This results in a shorter extension of the neutron detection plateau due to an approximation of its plateau knee, to the one of MIPs.

Pulse-height spectra of the cathode signals were recorded at 2100 V for three conditions: (1) the beam is on; (2) the beam is on, but a 5 mm thick $^{10}B_4C$ plate is placed in front of the detector window; (3) the beam is off. The results are shown in Figure 7.

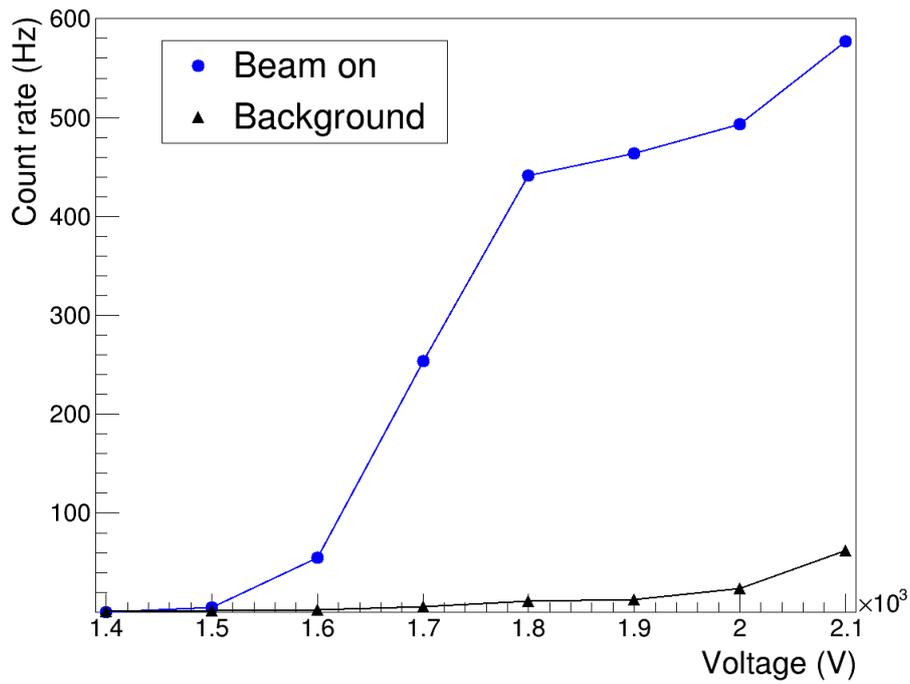

Figure 6. Measured count rate as a function of the negative voltage applied to the cathode. Blue circular dots: the detection module is irradiated by a neutron beam at the NARZISS beamline. Square black dots: background data (the beam is off). The lines are only an eye guide.

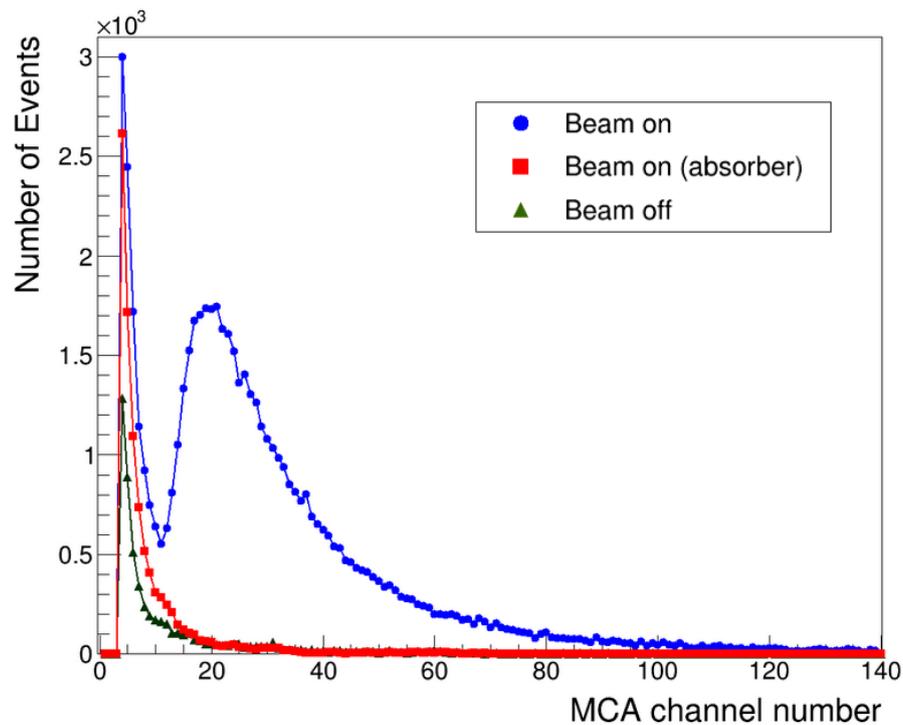

Figure 7. Measured charge spectra: the beam is on (round dots); the beam is on, but a $B_4C$ neutron absorber is placed in front of the detector window (square dots); the beam is off (triangular dots). All spectra were acquired during the same time interval of 100 s. There are no structures present outside the shown region.

The spectrum obtained in condition (1) includes contributions from the neutron events, the dark counts of the RPC, the background caused by the ambient radiation as well as the gamma rays originating from the interaction of the beam with the monochromator and the slits. For condition (2), the spectrum is contributed by the dark counts, the background caused by the ambient radiation, and gamma rays originating from the beam interactions (including capture in the absorber). Finally, for condition (3) the contributions are limited to the dark counts and ambient radiation events.

The beam-on curve (1) in Figure 7 shows a deep valley with a minimum close to channel number 10. Comparing the curves, one can conclude that neutrons give the dominating contribution above this channel number. Also, setting the detection threshold at the value corresponding to this channel will not result in a significant loss of neutron events.

The shape of the spectra, in particular the absence of saturation, also indicates that the RPC is operating in avalanche mode without the presence of streamers, thus validating the design of the detection module and its assembling procedure.

### 3.2. Neutron Time of Flight

For each detected neutron, its ToF is determined by calculating the time difference between two timestamps: one provided by the chopper and one defined by the fast component of the cathode signal. Figure 8 shows the distribution of the number of neutron events measured as a function of the neutron ToF.

To validate the obtained timing data, we have computed the wavelength spectrum of the beam at the BOA beamline which can be compared with literature data. This computation involves applying two procedures for all data points appearing in Figure 8: 1) conversion from the ToF to the neutron wavelength and 2) dividing the number of detected neutrons by the relative detection efficiency at the corresponding wavelength (chapter 2.4 and Figure 5). The obtained spectrum is shown in Figure 9. It closely resembles the data appearing in the review dedicated to the neutron imaging options at the BOA beamline [15].

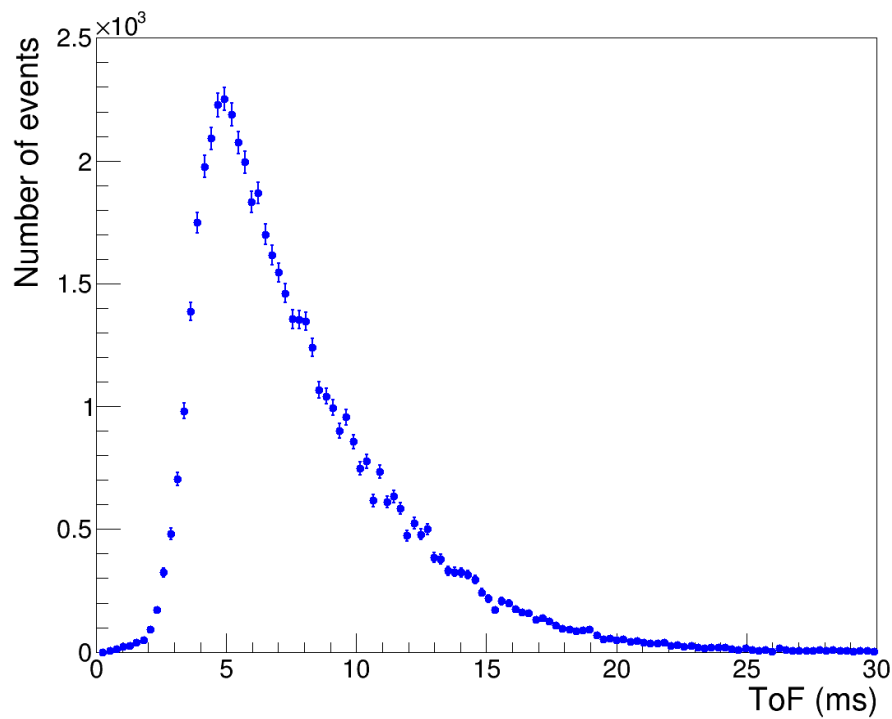

Figure 8. Distribution of the number of neutron events as a function of the time of flight.

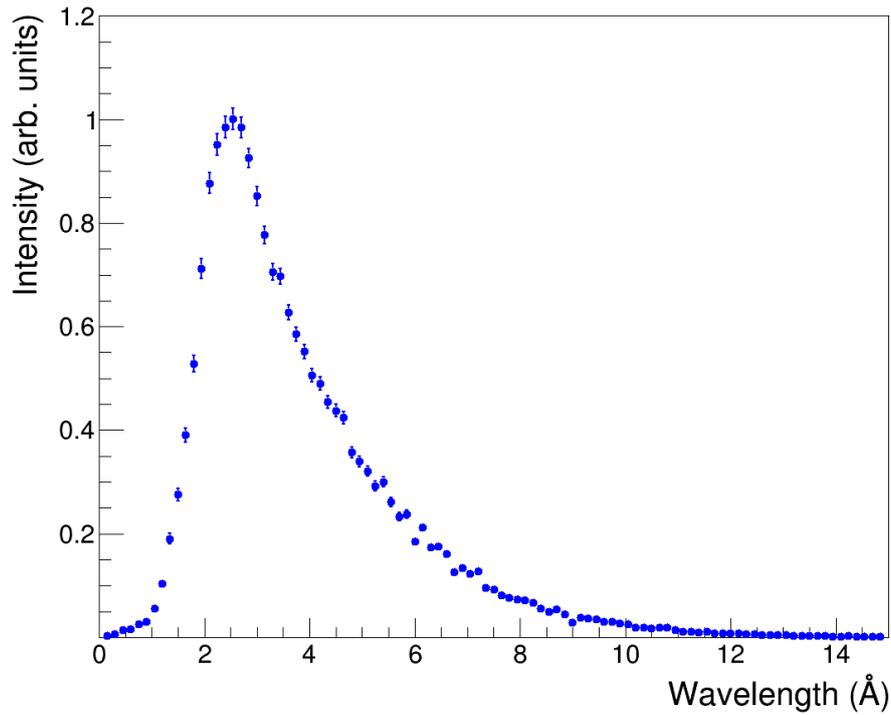

Figure 9. The computed wavelength spectrum of the beam at the BOA beamline. The data correspond to the beam centre area of 10 x 10 mm$^2$.

### 3.3. Determination of the triggered gas gap

One of the objectives of this study is to experimentally verify that the ratio of the sum signals for X and Y directions can be used to identify the $^{10}B_4C$ converter layer where the neutron was captured (see chapter 2.1.2). The sum of the amplitudes of the X and Y strip signals was calculated for each event in a dataset recorded at the BOA beamline and the obtained distributions are shown in Figure 10.

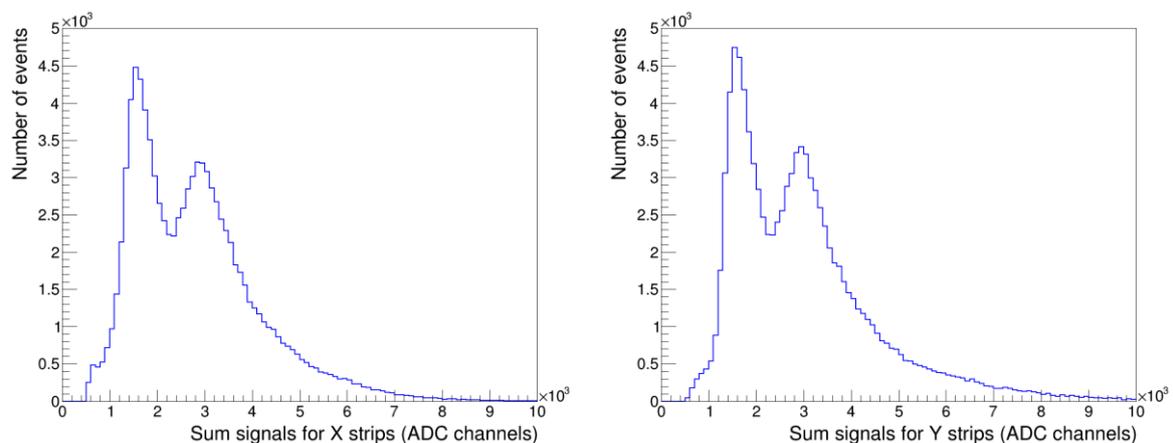

Figure 10. Histograms of the sum of the amplitudes of the X (left) and Y strip signals (right). Both distributions exhibit two peaks, which is explained by the presence of two types of events corresponding to the neutron capture in the upper or the lower $^{10}B_4C$ layer of the cathode.

Each distribution exhibits two peaks, both of a comparable area. These peaks can be explained by the presence of two types of events corresponding to the neutron capture in either the upper or the lower $B_4C$ layer of the same cathode. As the upper gas gap is faced by the X strips (see Figure 3), the Y strips are partially screened by the X ones, and, therefore, the sum of the X signals is larger than the Y sum for the neutrons captured in the upper converter layer. On the contrary, the lower gas gap is faced by the Y strips, so for neutrons captured in the lower converter the sum signal of Y strips is larger than that for X strips. As neutron events split nearly equally between the two converters, the dataset has equal fractions of both types of events.

To make the presence of two types of events even more clear, Figure 11 shows a heatmap of the event density versus the sum signal in the X and Y directions. One can see that there are two well-separated groups of events.

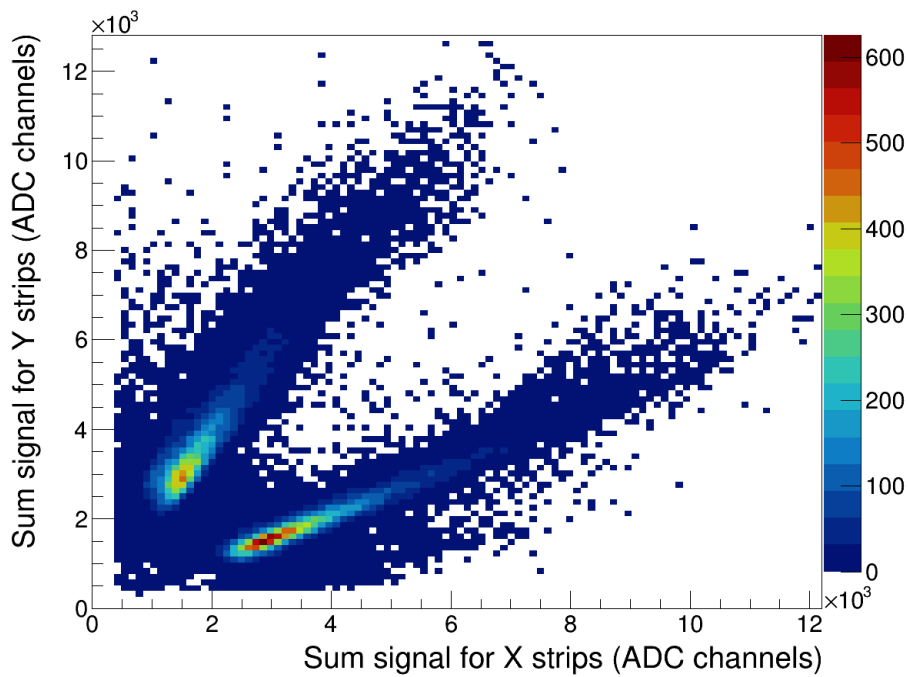

Figure 11. Heatmap of the event density versus the sum signal in X and Y directions.

The observed good separation of the two types of events allows us to conclude that the selected design of the XY position readout unit is adequate for identifying the $^{10}B_4C$ layer where the detected neutron was captured. This approach is especially useful for detectors with a large stack of detection modules, where only the cathode signal is used to identify the triggered cathode leaving ambiguity in the converter layer. Elimination of this ambiguity allows the improvement of two aspects of the detector performance. First, the accuracy of the neutron ToF determination is enhanced by pinpointing the neutron conversion point along the beam direction, eliminating the uncertainty of about 0.4 μs corresponding to the neutron flight time through the cathode plate. Second, the XY spatial resolution of the detector can be improved by taking into account the fraction of a millimetre misalignments between the arrays of the upper and lower interconnected strips, often introduced during the detector assembly [5].

## 4. Conclusions

In this study, we have experimentally demonstrated the feasibility of the new neutron RPC detection module and the position readout units of the new design. The module features a narrower gas gap (0.28 mm in comparison with the previous model with 0.35 mm), which resulted in the plateau appearing at significantly lower voltages (1800 V instead of 2050 V) but also extending for a shorter range (200 V compared to 500 V).

We have demonstrated that the new position readout unit makes it possible to identify in which of the two $^{10}B_4C$ converter layers of the same cathode the neutron was captured based on the ratio of the sum signals in the X and Y readout strips.

For the first time, a position-sensitive thermal neutron detector based on the $^{10}$B-RPC detection technology was used to measure the neutron time of flight. The wavelength spectrum of the BOA beamline at PSI, computed based on the measured ToF data, shows a good match with the literature data.

Currently, we are working on the optimization of a recently assembled detector prototype comprising a stack of ten neutron detection modules and eleven readout units. A series of tests to characterise the detector performance is planned for 2024.

## Acknowledgments

This work was supported by Portuguese national funds OE and FCT-Portugal (DOI 10.54499/EXPL/FIS-NUC/0538/2021) and is also based on experiments performed at the Swiss spallation neutron source, SINQ, Paul Scherrer Institute, Villigen, Switzerland. We would like to acknowledge the support of the Albert Einstein Center for Fundamental Physics (AEC) visitor programme at the University of Bern.